\documentclass[aps,pre,groupedaddress,twocolumn]{revtex4-2}

\usepackage{amsmath}    
\usepackage{graphicx}   
\usepackage{verbatim}   
\usepackage{color}      
\usepackage{subfigure}  
\usepackage{hyperref}   
\newcommand{\bq}{\begin{equation}}
\newcommand{\eq}{\end{equation}}
\newcommand{\bqn}{\begin{eqnarray}}
\newcommand{\eqn}{\end{eqnarray}}
\newcommand{\nb}{\nonumber}
\newcommand{\lb}{\label}

\newcommand{\f}{_{_{\text{freeze}}}}

\begin{document}


\title{Estimation of the Mass of Dark Matter Using the Observed Mass Profiles of Late-Type Galaxies}


\author{Ahmad Borzou}
\email[]{ahmad\_borzou@baylor.edu}
\affiliation{EUCOS-CASPER, Department of Physics, Baylor University, Waco, TX 76798, USA}


\date{\today}

\begin{abstract}
The system of stability equations for galactic halos is under-determined in most of the models of dark matter (DM).  
Conventionally,  the issue is resolved by taking the temperature as a constant,  and the chemical potential and the mass density as position-dependent variables.   
In this paper,  to close the under-determined set of equations,  we remove the mass density using observations and leave the temperature and the chemical potential as position-dependent variables.  
We analyze observations of the mass profiles of 175 late-type galaxies in the Spitzer Photometry \& Accurate Rotation Curves (SPARC) database as well as 26 late-type dwarfs in the Little Things database,  to construct the temperature profile of their DM halos by assuming that (1) DM in the halos obeys either the Fermi-Dirac or the Maxwell-Boltzmann distribution, and (2) the halos are in the virial state. 
We derive the dispersion velocity of DM at the center of the halos and show that its correlation with the halo's total mass is the same as the one estimated in N-body simulations and consistent with the direct observations of visible matter. 
Taking the latter agreement as a validation of our analysis, we derive the mass to the temperature of DM at the edge of the halos and show that it is galaxy independent and is equal to $m/T_{R_{200}}\simeq 10^{10}$ in natural units.  In the thermal models of DM,  such universal temperature is inherently assumed.  In this paper,  we derive the universal temperature from observations without imposing it by assumptions. 
Therefore,  $T_{R_{200}}$ in the above ratio can be expressed in terms of the temperature of the cosmic microwave background (CMB) at the time of DM decoupling. 
This result is used to study possible cosmological scenarios.
We show that observations are at odds with (1) non-thermal DM, (2) hot DM, and (3) collision-less cold DM.
If DM is warm,  we estimate its mass to be in the range of keV--MeV. 
\end{abstract}


\maketitle

\section{Introduction}
\lb{Sec:intro}
Various astronomical and cosmological observations point out that a mysterious DM constitutes about 85\% of the mass of the universe. A review of the pieces of evidence for the existence of DM can be found in \cite{2017NatAs...1E..57P}. Despite the large stockpile of data that refer to the presence of DM, only little is known about the nature of the dark particles, and widely different DM schemes are being investigated at the present. 

Cold collision-less dark matter (CDM) is among the most popular models.  Nonetheless,  N-body simulations of CDM result in singular density profiles,  cusps,  at the center of halos \cite{1996ApJ...462..563N,1997ApJ...490..493N} while observations  are in favor of core density profiles \cite{2009EAS....36..133S,  2011ApJ...742...20W,  2007ApJ...663..948G}.  
Also,  CDM may be the origin of several other discrepancies with observations of scales smaller than $\sim1\,$kilo-parsecs \cite{1999ApJ...522...82K,1999ApJ...524L..19M, 1991ApJ...378..496D, 1996ApJ...462..563N,2011MNRAS.415L..40B,2012MNRAS.422.1203B}.  
A proposed solution to the core-cusp problem is feedback from visible matter \cite{1996MNRAS.283L..72N,  1999MNRAS.303..321G}.  
An alternative popular solution to the core-cusp problem is a warm fermionic dark matter (WDM) where DM cannot form cusps due to the Pauli blocking \cite{2013NewA...22...39D, 2015JCAP...01..002D}.  To distinguish between the two mentioned proposals,  in this paper,  we evaluate the two scenarios at the edge of halos where visible matter's contribution is negligible.  We start with a framework that does not prioritize any of CDM or WDM by assumption.  Adding data to the framework favors WDM and disfavors CDM.

Among a few characteristics of DM that are known with high certainty is its galactic mass density, which is estimated through observations of how visible matter moves in galaxies. Observations also suggest that DM halos in galaxies are stable. Therefore, we can assume with high certainty that DM has a statistical distribution in the galactic halos, which leads to a pressure that confronts the attractive gravitational force. See \cite{2019A&ARv..27....2S} for a review.  Depending on its spin, DM distribution in galaxies is either Bose-Einstein or Fermi-Dirac, both of which are equivalent to the Maxwell-Boltzmann distribution under certain classical conditions. Since the visible mass in the universe is made of fermions and DM contributes to the mass of the universe as well, it is more likely that DM similarly obeys a Fermi-Dirac distribution. Finally, the virial theorem suggests with high certainty that DM halos that have not been involved in massive collisions, i.e. major mergers, are in the virial state.  

In this paper,  unlike the conventional approaches,  we do not make any assumptions about the temperature profile and the chemical potential profile of DM.  Instead,  we use the above-mentioned properties of DM to estimate its temperature profile in galaxies. We show that the temperature of DM at the edge of more than 200 observed late-type galaxies is approximately the same, and derive it in terms of the mass of DM. The analyzed galaxies are up to 100 mega-parsecs away from each other and have a diverse range of total masses from $10^8$ to $10^{13}$ in the unit of the mass of the sun. We conclude that DM is a thermal relic and its universal temperature at the edge of the halos is equal to its temperature when the dark particles decoupled from the rest of the matter in the early universe and cooled down due to the expansion of the universe.  While the latter result is the natural consequence in all thermal models of DM,  it is not assumed but derived in this paper.
As a validation of our analysis, we estimate the temperature to the mass of DM at the center of the observed halos and show that it is the same as the observed temperature to the mass of visible matter, and also consistent with N-body simulations. 

We investigate the implications of the estimated universal DM temperature within the context of a few popular cosmological models.
While CDM is not consistent with our findings, we estimate the mass of WDM to be in the interval of keV--MeV. 

This paper is structured as follows. 
In sections~\ref{Sec:Stability}, and~\ref{Sec:Virial}, we lay the theoretical framework to express the temperature of DM in terms of its observable mass density. In section~\ref{sec:Sig2OfSPARC}, we estimate the temperature profile of more than 200 observed late-type galaxies and show the universality of the temperature of DM at the edge of their halos. 
We study the effects of visible matter in section~\ref{Sec:VisibleMatter}.
The implications of the universal temperature for different DM models are investigated in section~\ref{sec:Cosmology}. 
A conclusion is drawn in section~\ref{sec:conclusion}.

\section{Stability of halos}
\lb{Sec:Stability}
To maintain the stability of the halos, the forces of gravity and pressure should be equal at any distance from the center
\bqn
\lb{Eq:Stability}
-\frac{1}{\rho(r)}\frac{dP}{dr} = G\frac{M(r)}{r^2}, 
\eqn
where $\rho(r)$ is the mass density of DM, $P(r)$ is the pressure of DM, $G$ is the Newton gravitational constant, and $M(r)$ is the DM mass enclosed in radius $r$.  DM halo is assumed spherically symmetric,  as suggested by observations \cite{1969ApJ...155..393P,2001ApJ...555..240B,2006ApJ...638L..13T}.

In this paper,  our strategy is (i) to build a general framework that does not prioritize any of CDM or WDM,  and (ii) to insert data into the framework to evaluate the DM models.  
Fermi-Dirac statistics meets our purposes since it naturally describes fermionic WDM,   while,  depending on the mass and temperature of DM,   reduces to a Maxwell-Boltzmann distribution that describes CDM and even heavy bosons.  In \cite{2020EPJC...80.1076B},  through various examples,  it is shown that Fermi-Dirac statistics can reproduce the classical solutions that are often attributed to CDM.  

Therefore, 
the mass density,  and the pressure in a wide range of DM models are given by the Fermi-Dirac statistics through
\bqn
\lb{Eq:P_rho}
&&\rho(r) = \frac{2m}{\alpha^3}\left(kT(r)\right)^{\frac{3}{2}}f_{\frac{3}{2}}(r),\nb\\
&&P(r) = \frac{2}{\alpha^3} \left(kT(r)\right)^{\frac{5}{2}}f_{\frac{5}{2}}(r),
\eqn
where $m$ being the mass of DM,  $k$ is the Boltzmann constant, $T(r)$ is the temperature of DM,
and $\alpha$ is a constant in terms of the mass of DM and the Planck constant. 
Also, the Fermi-Dirac integrals are defined as 
\bqn
f_{\nu}(z(r))=\frac{1}{\Gamma(\nu)}\int_0^\infty\frac{x^{\nu-1}dx}{z(r)^{-1}e^x+1},
\eqn
where the gamma function is shown with $\Gamma(\nu)$,  and the fugacity is defined in terms of the chemical potential $\mu(r)$ as $z(r) = \exp(\mu(r)/kT(r))$.

Using equation~\eqref{Eq:P_rho},  the pressure of DM reads
\bqn
\lb{Eq:EOS}
P(r) = \sigma_0^2 \, y(r) \rho(r),
\eqn
where the subscript naught refers to the value of the quantities at the center,  and
\bqn
&& \sigma^2(r)=k T(r) h(r)/m,\lb{Eq:Sig2}\\
&& y(r) \equiv \sigma^2(r)/\sigma^2_0, \lb{Eq:y_sig2}\\
&& h(r)\equiv f_{\frac{5}{2}}(r)/f_{\frac{3}{2}}(r).\lb{Eq:h}
\eqn
Here,  the first line refers to the dispersion velocity squared using the Fermi-Dirac statistics.  The second line is the dimensionless dispersion velocity squared.  The third line determines whether the distribution has reduced to the Maxwell-Boltzmann form and $h(r)=1$,  or the quantum nature of DM is significant and $h(r) > 1$.

Before moving forward,  we need to clarify a few points.  
First,  the system of equations above is written for a free-falling observer.  The equations can be easily converted to a frame that is attached to the center of the halo through a simple transformation.  The two sets of equations return identical solutions as is shown in \cite{2020EPJC...80.1076B}. 
Second,  the system of equations above is under-determined.  There exist four unknown variables to be solved,  namely 
$\left[\rho(r),  T(r),  \mu(r), P(r) \right]$.  On the other hand,  we only have three equations to solve,  one equation in~\eqref{Eq:Stability},  and two equations in~\eqref{Eq:P_rho}.  Therefore,  we need to remove one of the variables by assumption.  

One popular approach is to set $T(r)=T_0$.  For example see \cite{2014MNRAS.442.2717D, deVega:2013woa}.  
In an interesting alternative,  presented in \cite{2016IJMPA..3150073D},  the mass density $\rho$ is assumed to be equal to one of the phenomenological mass models,  but the distribution function $f(r)$ is a variable to be solved for.  Therefore,  the list of unknown variables is changed to $\left[f(r),  T(r),  \mu(r), P(r) \right]$.  It should be noted that the phenomenological mass density serves as the exact, not approximate,  solution.  Still,  however,  the system of equations is not closed.  The authors assume a constant temperature $T(r)=T_0$ to close the equations.  The latter can be understood from the form of the distribution function that they impose,  $f(r)=f(p^2/2m - \mu(r))$ with $p$ being the momenta of DM particles.
Nevertheless,  the authors refer to the dispersion velocity squared as an effective temperature.  We emphasize that their definition of effective temperature is different from ours in this article.  We define temperature as the parameter $T(r)$ that enters the Fermi-Dirac distribution in the following way
\bqn
\lb{Eq:FermiDiracDist}
f(r) = \Bigg[1+\exp\bigg( \Big(kT(r)\Big)^{-1}\cdot\Big(p^2/2m - \mu(r)\Big) \bigg)\Bigg]^{-1},\nb\\
\eqn
which leads to equation~\eqref{Eq:P_rho}.
In \cite{2020EPJC...80.1076B},  a software is introduced that assumes this distribution and finds the solutions for any assumed form of $T(r)$.  The conventional $T(r)=T_0$ is compared with a generic profile of the form $T(r)=T_0[1+r^2]^{-1}$.  Notably,  warmer halos are larger in the former but smaller in the latter.  

In this paper,  we start with the Fermi-Dirac distribution in equation~\eqref{Eq:FermiDiracDist}.  We close the system of equations by assuming that the phenomenological mass models are exact solutions for $\rho(r)$.  That means,  we solve the three equations in~(\ref{Eq:Stability},~\ref{Eq:P_rho}) to find the following three unknown variables, $\left[T(r),  \mu(r), P(r) \right]$.  
Using equations~\eqref{Eq:Stability} and \eqref{Eq:EOS}
\bqn
\lb{Eq:y_r}
y(r) = \frac{\rho_0}{\rho(r)}\left(1-\frac{G}{\rho_0\sigma_0^2}\int_0^{r}\frac{\rho(r')M(r')}{r'^{2}}dr'\right).
\eqn
It should be noted that the equation above is valid regardless of the degeneracy level of fermionic DM. Even when the temperature tends to zero, corresponding with the full degeneracy level,  $h(r)$ tends to infinity such that the right-hand side remains finite and non-zero. The equation is also valid for CDM and heavy bosonic DM in which case $h(r)=1$.  It should be noted that we do not assume any value or form for $h(r)$ to avoid prioritizing DM models.  Instead,  we intend to insert data into the right-hand side of equation~\eqref{Eq:y_r} to find $y(r)$, which subsequently will be decomposed into its $T(r)$ and $h(r)$ components.  In other words,  we will let data prioritize a DM model.

In general, $\rho(r)$ and $\sigma_0^2$ in the equation above are independent and should be separately derived from observations. Nevertheless, by observing the mass profile, we can estimate a lower bound on the dispersion velocity. 
Since the left-hand side of equation~\eqref{Eq:y_r} is positive by definition, the central dispersion velocity has to satisfy the following inequality \cite{2020EPJC...80.1076B}
\bqn
\lb{Eq:Sig2LowerBound}
\sigma_0^2 > \frac{G}{\rho_0} \text{Max}\left(\int_0^{r}\frac{\rho(r')M(r')}{r^{'2}}dr'\right),
\eqn
where 'Max' refers to the value of the integral at a distance $r$ where the integral reaches its maximum.  

The inequality in equation~\eqref{Eq:Sig2LowerBound} is complimentary to the  inequality attributed to Tremaine \& Gunn.  If DM has a fermionic nature,  its dispersion velocity has a minimum due to the limitation of the phase-space of fermions.  
Since (i) we are starting with the Fermi-Dirac statistics,  and (ii) the system of equations is under-determined,  the Tremaine-Gunn inequality is naturally respected in all of our solutions \cite{2020EPJC...80.1076B}.  Still,  we can easily propose a set of $\left[\sigma_0, \rho(r)\right]$ that violates the inequality in equation~\eqref{Eq:Sig2LowerBound}.  In this paper, we show that,  miraculously,  all of the observed galaxies have a set of $\left[\sigma_0, \rho(r)\right]$ consistent with the latter inequality.

\section{Virial State}
\lb{Sec:Virial}
In this section, we use the virial theorem to derive the dispersion velocity of DM at the center of the halos in terms of $\rho(r)$, such that $y(r)$ in equation~\eqref{Eq:y_r} is entirely known if the mass density is constructed out of observations. 

According to the virial theorem, the total kinetic energy $U$ of DM halo is equal to minus half of its total gravitational potential energy $W$ if the galaxy has not been participating in a major merger recently.

The gravitational potential energy of the halo is given by
\bqn
\lb{Eq:GravityPotentialEnergy}
W = \frac{1}{2}\int_0^{R} 4\pi r^2  \rho(r)\phi(r) dr, 
\eqn
where the gravitational potential is 
\bqn
\lb{Eq:GravityPotential}
\phi(r)=-4\pi G\left(   \frac{1}{r}\int_0^r \rho(r')r'^{2}dr' + \int_r^R\rho(r')r'dr'  \right).
\eqn 

Since the density of the kinetic energy of DM is equal to $\frac{3}{2}P$, we use equations~\eqref{Eq:EOS} and \eqref{Eq:y_r} to calculate the total kinetic energy of the halos 
\bqn
\lb{Eq:TotalKineticEnergy}
U = 6\pi \rho_0\sigma_0^2 \int_0^R r^2 \left(1-\frac{G}{\rho_0\sigma_0^2}\int_0^{r}\frac{\rho(r')M(r')}{r'^{2}}dr'\right) 
dr.\nb\\
\eqn

Using the virial theorem, and equations~\eqref{Eq:GravityPotentialEnergy},~and~\eqref{Eq:TotalKineticEnergy}, the dispersion velocity of DM at the center of the halos read
\bqn
\lb{Eq:CentralDispersion}
\sigma_0^2 =  \frac{-1}{2\pi \rho_0 R^3}\left(\frac{1}{2}W+U_2\right),
\eqn
where $U_2$ refers to the second term on the right hand side in equation~\eqref{Eq:TotalKineticEnergy}.

Therefore, the ratio of the mass of DM $m$ over its temperature at the edge of the halo $T(R)$ reads
\bqn
\lb{Eq:DMmass}
\frac{m}{T(R)} = \frac{k}{\sigma_0^2 \, y(R)},
\eqn
where the right-hand side is known in terms of observed $\rho(r)$. Moreover,  
we have safely assumed that at the edge of the halos,  $h(R) \simeq 1$. 
In appendix~\ref{App:hR}, we specifically calculate  $h(R)$ for all of the reported scenarios and prove the validity of the latter assumption in them.

\section{Analysis of observed late-type galaxies}
\lb{sec:Sig2OfSPARC}
\begin{figure*}
\centering
\includegraphics[width=\columnwidth]{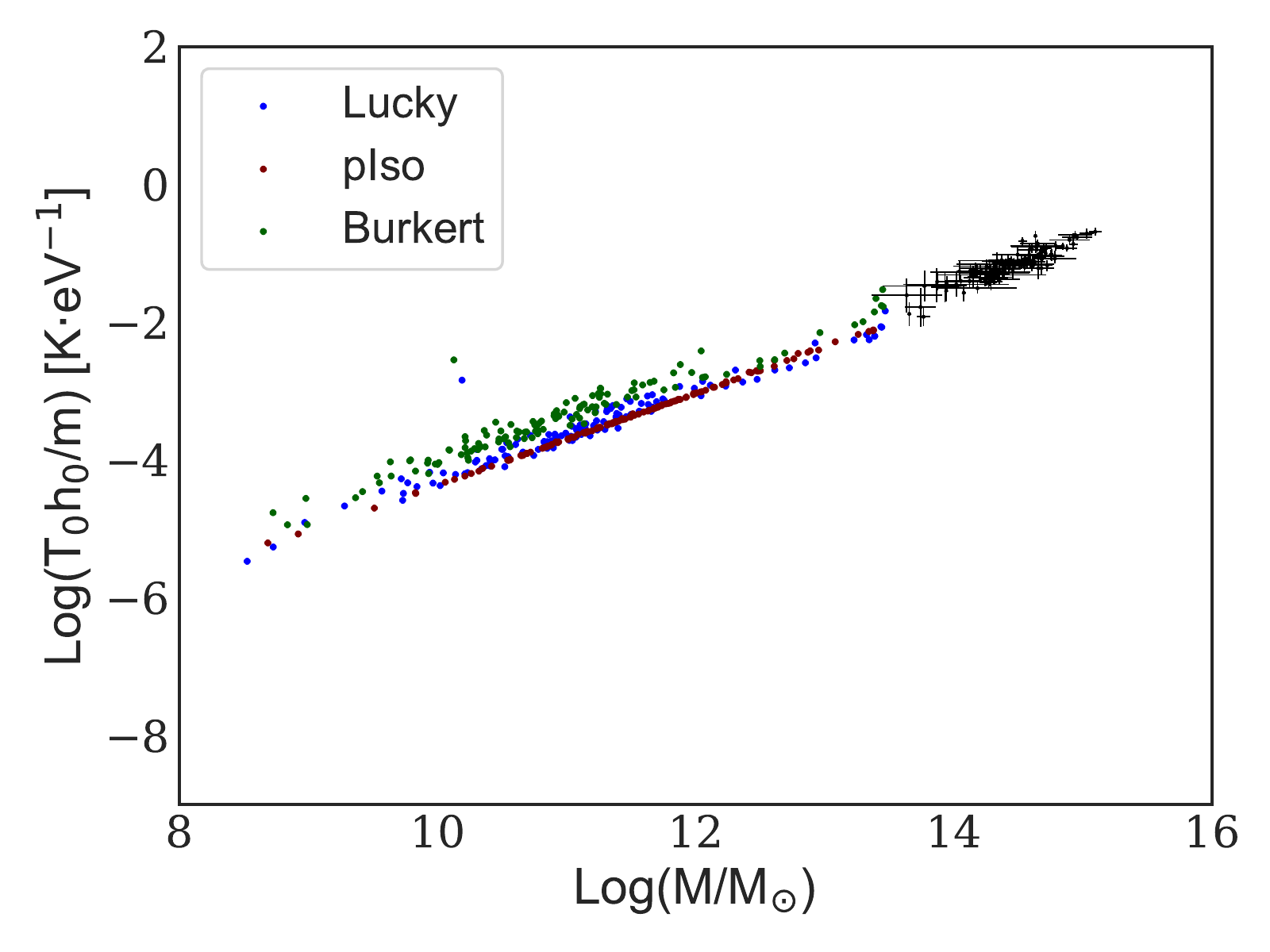}
\includegraphics[width=\columnwidth]{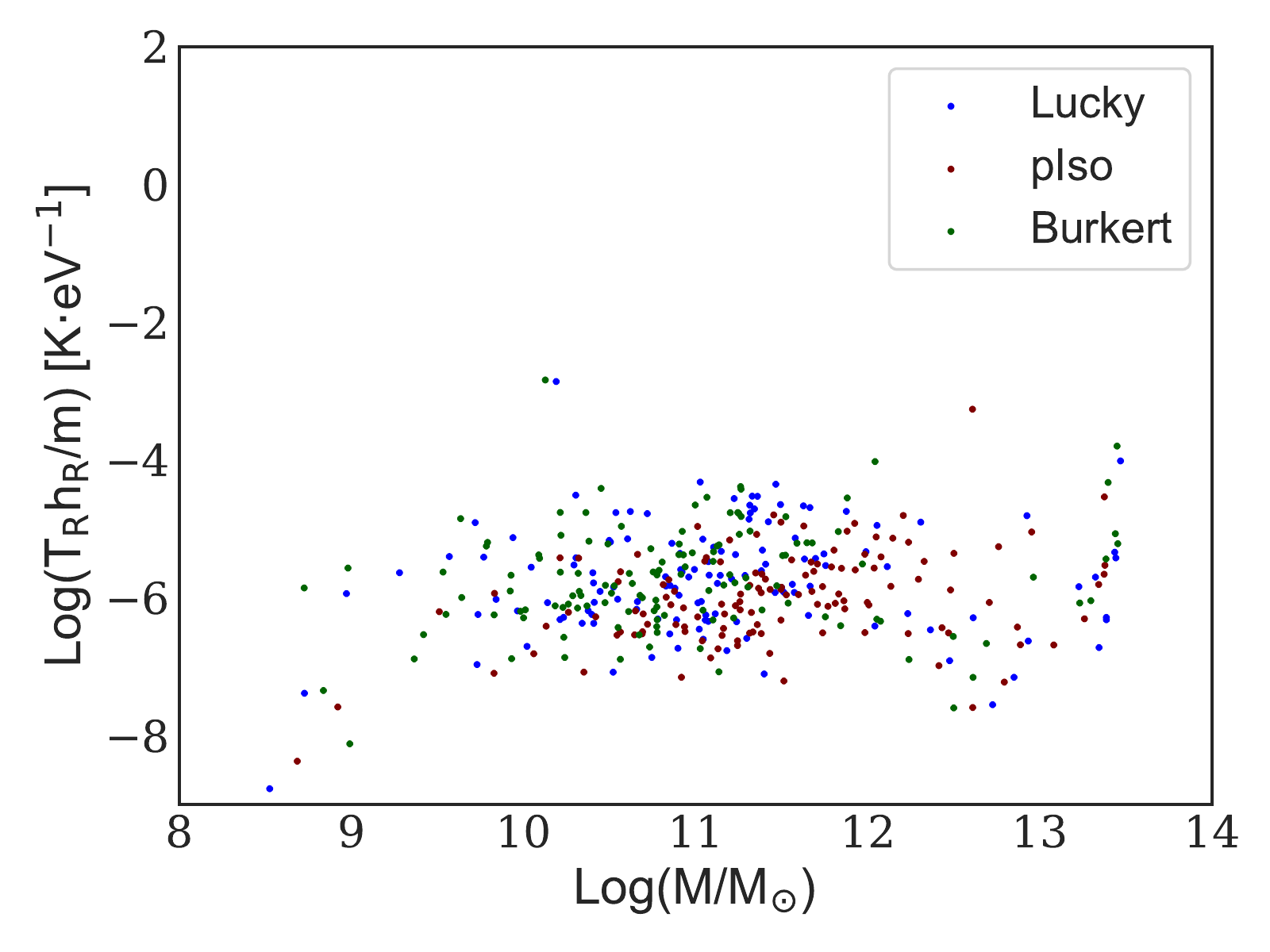}
\caption{The temperature to the mass of DM versus the total mass of the halos of the SPARC galaxies. The left panel refers to the DM temperature at the center of the halos.  The points with error bar in the left panel are from \cite{2016ApJ...819...63R} and show the observed ``temperature over mass'' of visible matter in Sunyaev-Zeldovich-selected clusters. This panel shows a remarkable agreement between dark matter and visible matter in terms of $T_0h_0/m-M_{200}$ relation. 
The right panel refers to the temperature of DM at the edge of the halos where the mass density of DM is only 200 times the critical mass density in the universe. 
It can be seen that the DM temperature at the edge of the halos is almost universal. The figure also shows that in light halos with the mass of $\sim 10^8\,$M$_{\odot}$, DM temperature at the center is almost the same as the universal outer temperature.  In thermal DM models,  we expect a universal relic DM temperature in unperturbed regions of the universe.  The right panel indicates such universal DM temperature even though we have not assumed but derived this universal temperature.  
\lb{Fig:T_m_M_R200rhoc}}
\end{figure*}
In this section, we use observations of 175 late-type galaxies in the SPARC database \cite{2016AJ....152..157L} and 26 late-type dwarfs in the Little Things database \cite{2015AJ....149..180O},  together with the theoretical framework presented in the preceding sections, to investigate the temperature of DM  in those galaxies.  

The SPARC database contains both H$_{\text{I}}/$H$_{\alpha}$ rotation curves and near-infrared surface photometry. The latter helps with the construction of the DM mass profile close to the center while the former can be used to learn the halos' outer mass. The observations are subsequently used to set the free parameters of a few popular mass models for each of the galaxies. 
The three mass models that we use in this paper are 
\bqn
\rho(r) = 
\begin{cases}
\rho_0\left[\left(1+\frac{r}{r_0}\right)\left(1+\left(\frac{r}{r_0}\right)^2\right)\right]^{-1} & \text{Burkert}\\
\rho_0\left[1+\left(\frac{r}{r_0}\right)^2\right]^{-1} & \text{pIso}\\
\rho_0\left[1+\left(\frac{r}{r_0}\right)^3\right]^{-1} & \text{Lucky13},\\
\end{cases}
\eqn
where,  for each of the 175 galaxies, the scale radius $r_0$ and the characteristic mass density $\rho_0$ are estimated using the observations and are provided in \cite{2020ApJS..247...31L}. 

We use the above three mass models as $\rho(r)$ in equations~\eqref{Eq:CentralDispersion} and \eqref{Eq:DMmass} to derive $\sigma^2_0$ and $T(R)/m$ of DM in the corresponding halos. We show that the two quantities do not strongly depend on the mass models of $\rho(r)$ as far as they are not singular at the center.

\begin{figure}
\centering
\includegraphics[width=\columnwidth]{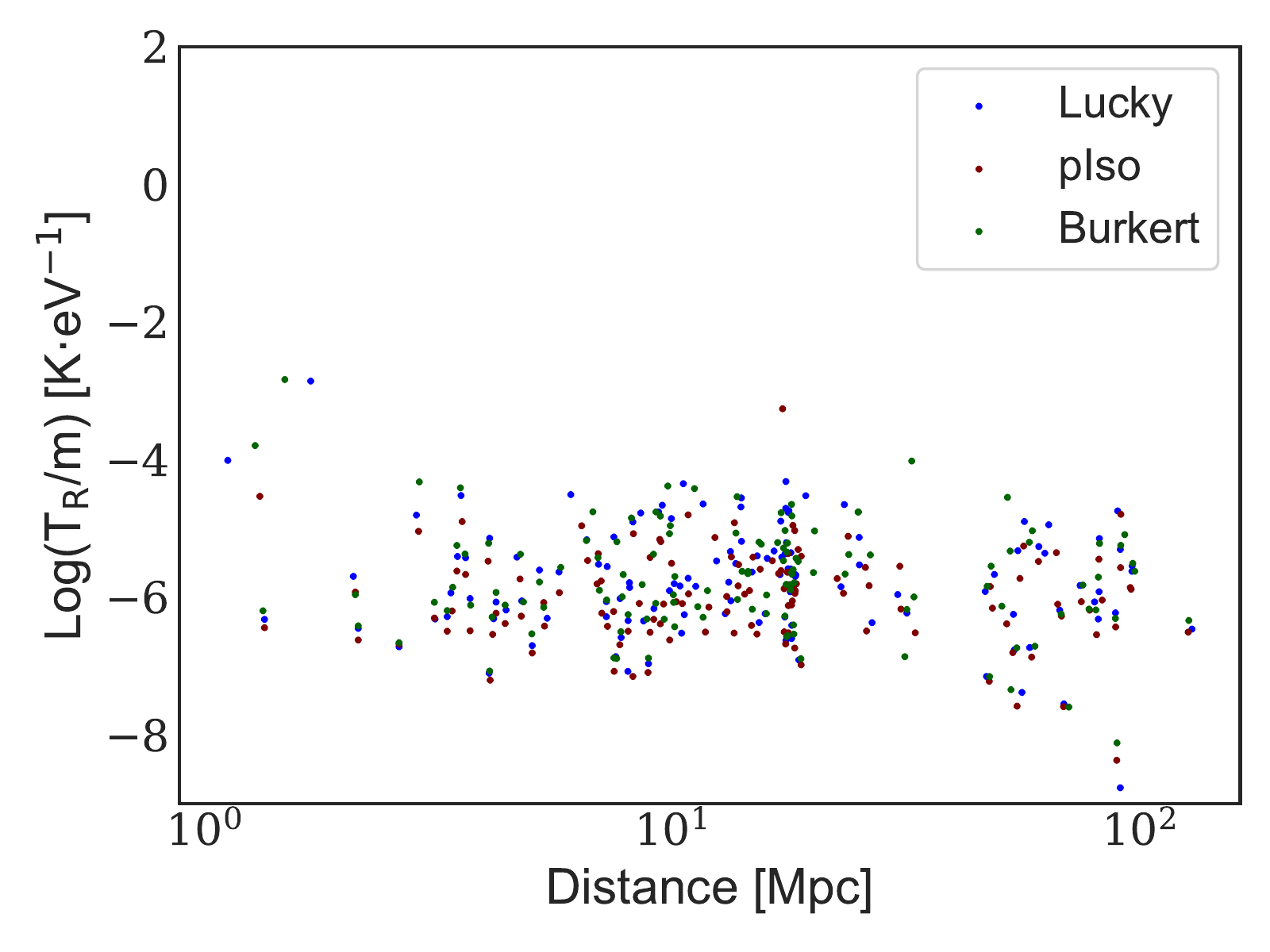}
\caption{The distance of the SPARC galaxies from us in the x-axis and the edge temperature over the mass of DM in the y-axis. The plot indicates that the halos are far enough from each other that cannot be in thermal equilibrium at the present epoch.  Yet,  the DM temperature at the edge of halos is universal indicating that the temperature is set in the early universe. \lb{Fig:TR_m_D}}
\end{figure}
Figure~\ref{Fig:T_m_M_R200rhoc} shows the temperature of DM divided by its mass for every galaxy in the SPARC dataset.   
The left panel shows that the temperature at the center depends on the total mass of the halos. A more conventional variant of this plot is shown in the appendix in figure~\ref{Fig:Sig0_M_R200rhoc}, which shows that the logarithm of the dispersion velocity of DM at the center of halos is linearly related to the logarithm of the total mass of the halos. The slope in this plot coincides with the same slope estimated in the N-body simulations of DM as well as with the observed slope for the dispersion velocity of visible matter \cite{2016ApJ...819...63R,2016ApJ...832..203Z}. This agreement is in favor of our assumption that the analyzed halos of the SPARC dataset are in the virial state. 
We report that the estimated dispersion velocities in this figure are approximately equal to the lower bound derived in equation~\eqref{Eq:Sig2LowerBound}.

The right panel of figure~\ref{Fig:T_m_M_R200rhoc} shows that the DM temperature at the edge of the halos is universal, and at 95\% confidence
\bqn
\lb{Eq:TR200_m}
\frac{T(R_{200})}{m}=\left(1.9 \pm 0.3\right)\times 10^{-6}\,(\text{K}\,\cdot\,\text{eV}^{-1}),
\eqn
where $R_{200}$ is conventionally defined as the edge of the halos, where the mass density of DM is 200 times the critical mass density.  
Figure~\ref{Fig:T_m_M_R200rhoc} also shows that in light halos of mass $\sim 10^8\,$M$_{\odot}$, the DM temperature at the center is not different from the outer universal temperature. This observation implies that, in such halos, the gravitational energy that is converted to the kinetic energy of dark particles is negligible. 
Figure~\ref{Fig:TR_m_D} shows that the temperature of DM at the edge of the halos is the same even though the galaxies are in a wide range of distances from us and cannot communicate as rapidly as needed to maintain a thermal equilibrium at the present.
Later in section~\ref{Sec:VisibleMatter},  we derive similar results for the SPARC galaxies after taking their visible matter into considerations.

\begin{figure*}
\centering
\includegraphics[width=\columnwidth]{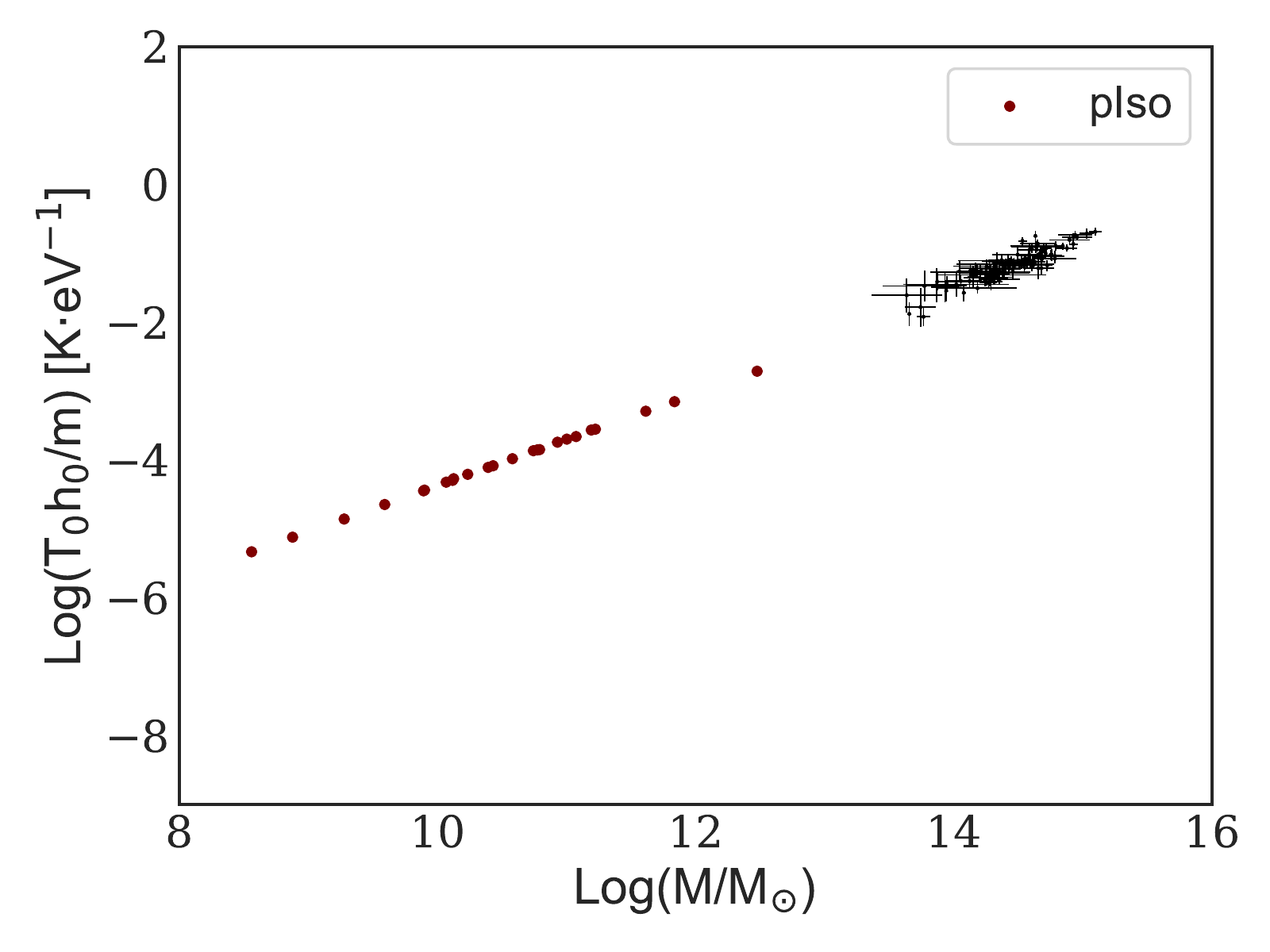}
\includegraphics[width=\columnwidth]{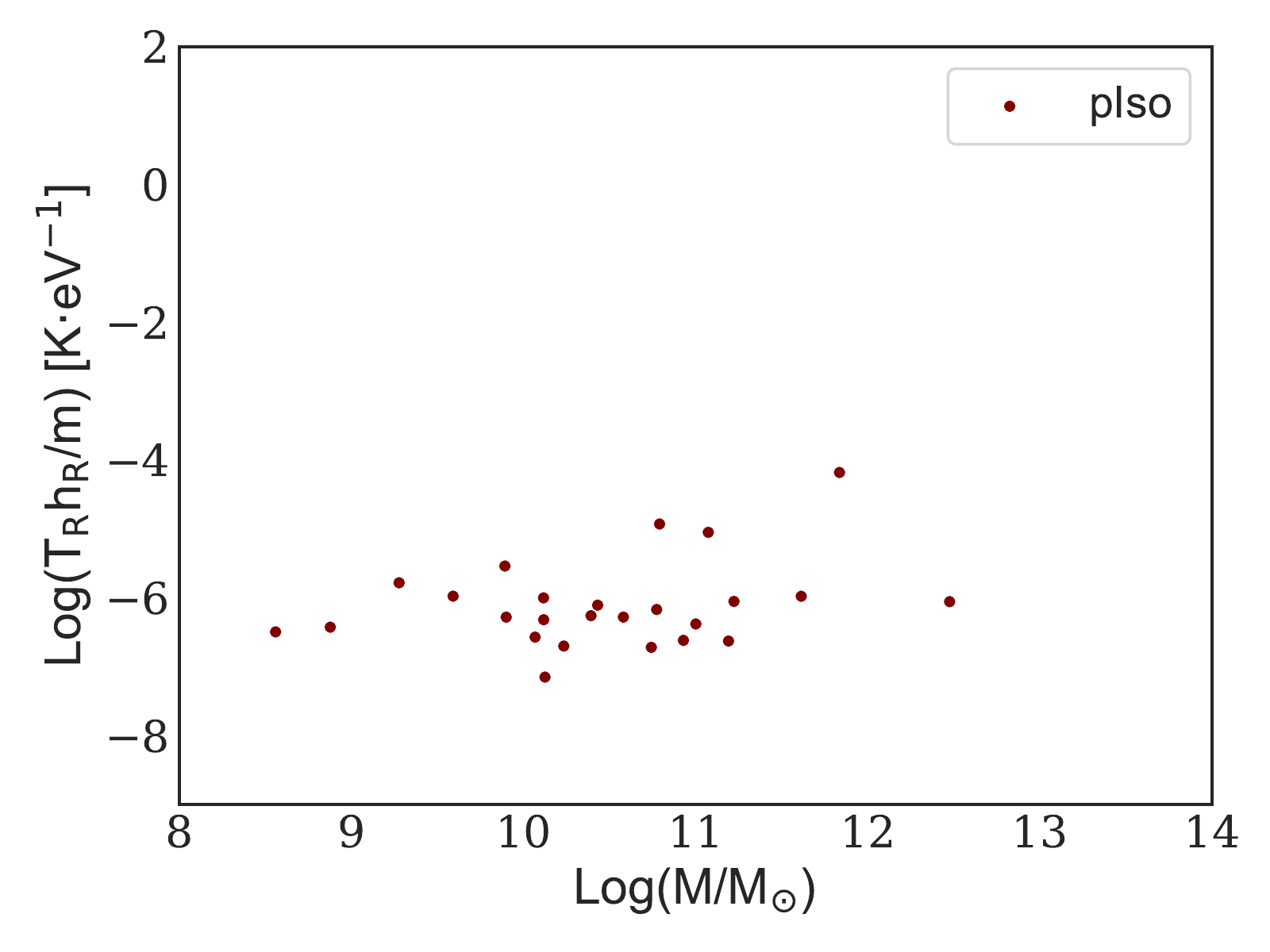}
\caption{The temperature to the mass of DM versus the total mass of the halos of the 26 late-type dwarf galaxies in the Little Things \cite{2015AJ....149..180O}. The left panel refers to the DM temperature at the center of the halos and the right panel refers to the temperature of DM at the edge of the halos where the mass density of DM is only 200 times the critical mass density in the universe. 
The points with error bar in the left panel are from \cite{2016ApJ...819...63R} and show the observed ``temperature over mass'' of visible matter in Sunyaev-Zeldovich-selected clusters.  The two panels are consistent with the corresponding ones in figure~\ref{Fig:T_m_M_R200rhoc}.   The temperature of DM at the edge of galaxies is universal suggesting that DM is a thermal relic.  Such universality is naturally assumed in the thermal models of DM.  Nonetheless,  we derive it from observations without imposing it by assumptions.  
Due to the absence of spiral and bulge components in these 26 dwarfs and due to the dominance of DM in their dynamics,  this plot provides a more reliable estimation of the temperature of DM. 
Note that the temperature in the right panel of this figure has less variance than in the right panel of figure~\ref{Fig:T_m_M_R200rhoc}.  The latter shows that DM is less perturbed at the edge of dwarfs with less contribution from visible matter.  
\lb{Fig:T_m_M_R200rhoc_LittleThings}}
\end{figure*}
Another approach to resolving the contribution of visible matter in our estimations of $\sigma_0^2$ and $T(R)/m$ would be to restrict ourselves to dwarf galaxies that do not have spiral and bulge components,  where DM has a dominant role in their dynamics.   In the following,  we repeat the procedure of this section and neglect the contribution from visible matter, for the 26 late-type dwarf galaxies from the local volume in the Little Things database.  Since only the pseudo-isothermal mass model is fitted to the observations in the latter reference,  we present the results for this mass model.  Nevertheless,  from figure~\ref{Fig:T_m_M_R200rhoc} it is already clear that our estimations for the temperature of DM are relatively independent of the mass model and similar results are expected for other mass models such as Burkert. 
Figure~\ref{Fig:T_m_M_R200rhoc_LittleThings} shows our estimations of DM temperature at the center and the edge of the halos of the dwarf galaxies in the Little Things.  The left panel establishes the remarkable agreement with the direct observations of visible matter.  The right panel shows that the temperature of DM at the edge of halos is universal.  Since the visible matter has a negligible contribution to the dynamics of dwarf galaxies,  and since such galaxies do not have spiral and bulge components,  the estimations of DM temperature in figure~\ref{Fig:T_m_M_R200rhoc_LittleThings} are more reliable than the ones from the SPARC dataset.  Notably,  the variance in the edge temperature of DM is reduced in the Little Things dwarfs.

The universality of DM temperature at the edge of the halos suggests that DM has been in thermal contact in the early universe. The decoupled DM keeps the distribution but cools down via the expansion of the universe.  Later on,  when the perturbations in the mass density of the universe start to grow,  the conversion of the gravitational potential warm DM up.  Therefore,  the temperature at the center of halos is higher than their relic temperature.  At the edge of the halos,  the temperature of DM is less perturbed and should be approximately equivalent to the relic temperature.    
Consequently, $T_R$ can be expressed in terms of the temperature of DM at the decoupling in the early universe, $T\f$.  It should be noted that similar to CMB and relic neutrino temperature,  the temperature of DM in thermal models is universal by assumption.  In this paper,  we do not assume such universal temperature but instead derive it from observations.

\begin{figure}
\centering
\includegraphics[width=\columnwidth]{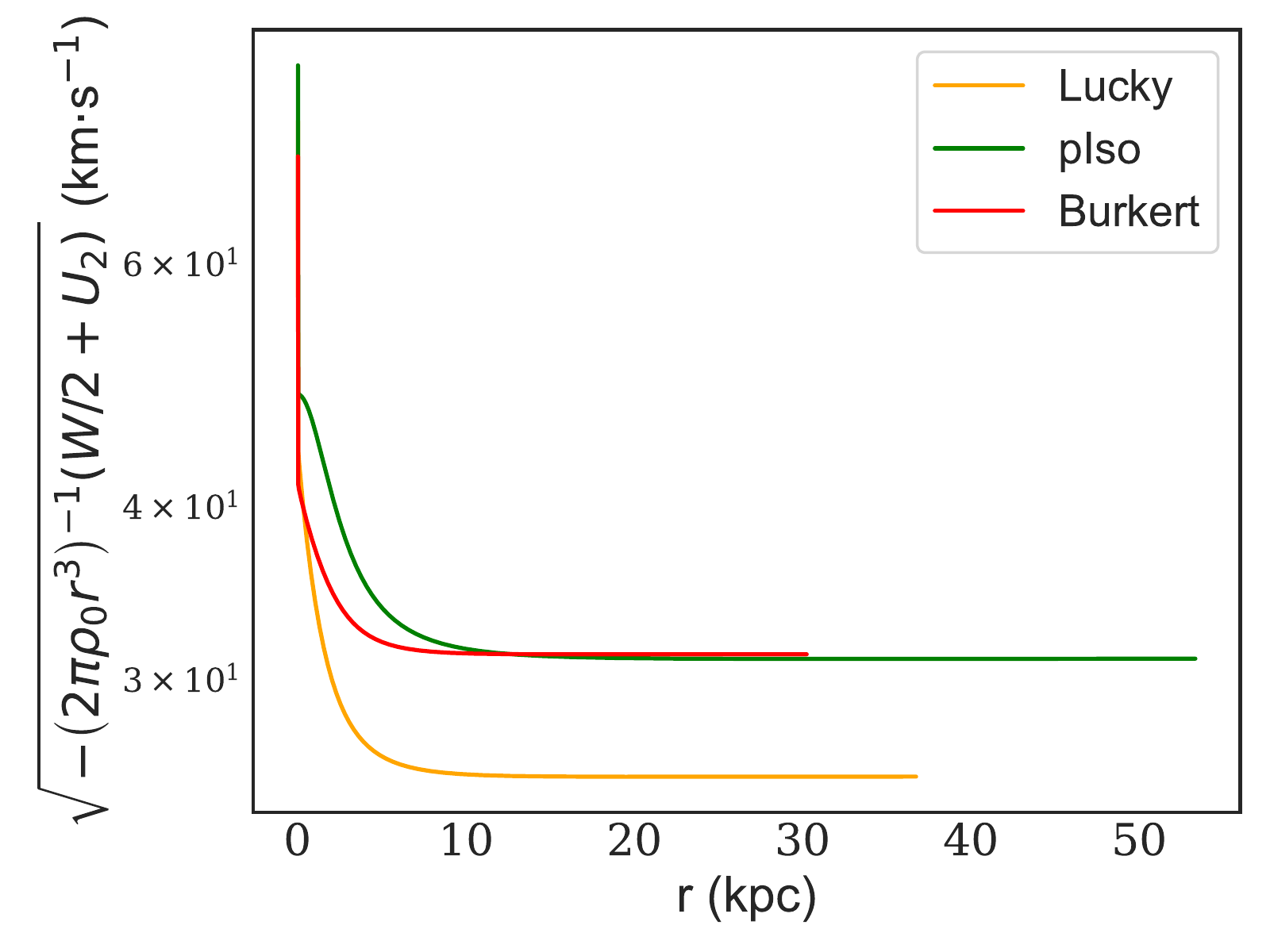}
\caption{The y-axis refers to equation~\eqref{Eq:CentralDispersion} with a varying halo radius. The three curves are cut on the x-axis where the density reaches as low as 200 times the critical density of the universe. As can be seen the y-axis reaches a flat plateau at small distances from the center, and does not depend on where $R$ is defined. The figure also indicates that $\sigma_0$ has a slight dependence on the mass model. However, with the order of magnitude precision, all of the mass models refer to $\sim 10$~(km $\cdot$ s$^{-1}$). We report the same behavior in all of the galaxies although this plot refers particularly to DDO170. \lb{Fig:Sig0IndependenceOfR}}
\end{figure}
Since the exact location where DM halo ends is not known with certainty and $R_{200}$ is only an estimation,
in the following, we investigate which one of our results depend on the definition of the edge of the halo, and quantify such possible dependencies. For the sake of ease in presentation, we only show the results for galaxy DDO170. 

To show that the dispersion velocity of DM at the center of halos is independent of the edge, we invoke equation~\eqref{Eq:CentralDispersion} and plot $\sigma_0$ in terms of $R$. The result can be seen in figure~\ref{Fig:Sig0IndependenceOfR} and indicates that after a small distance from the center, the right hand side of  equation~\eqref{Eq:CentralDispersion} reaches a constant value.

\begin{figure}
\centering
\includegraphics[width=\columnwidth]{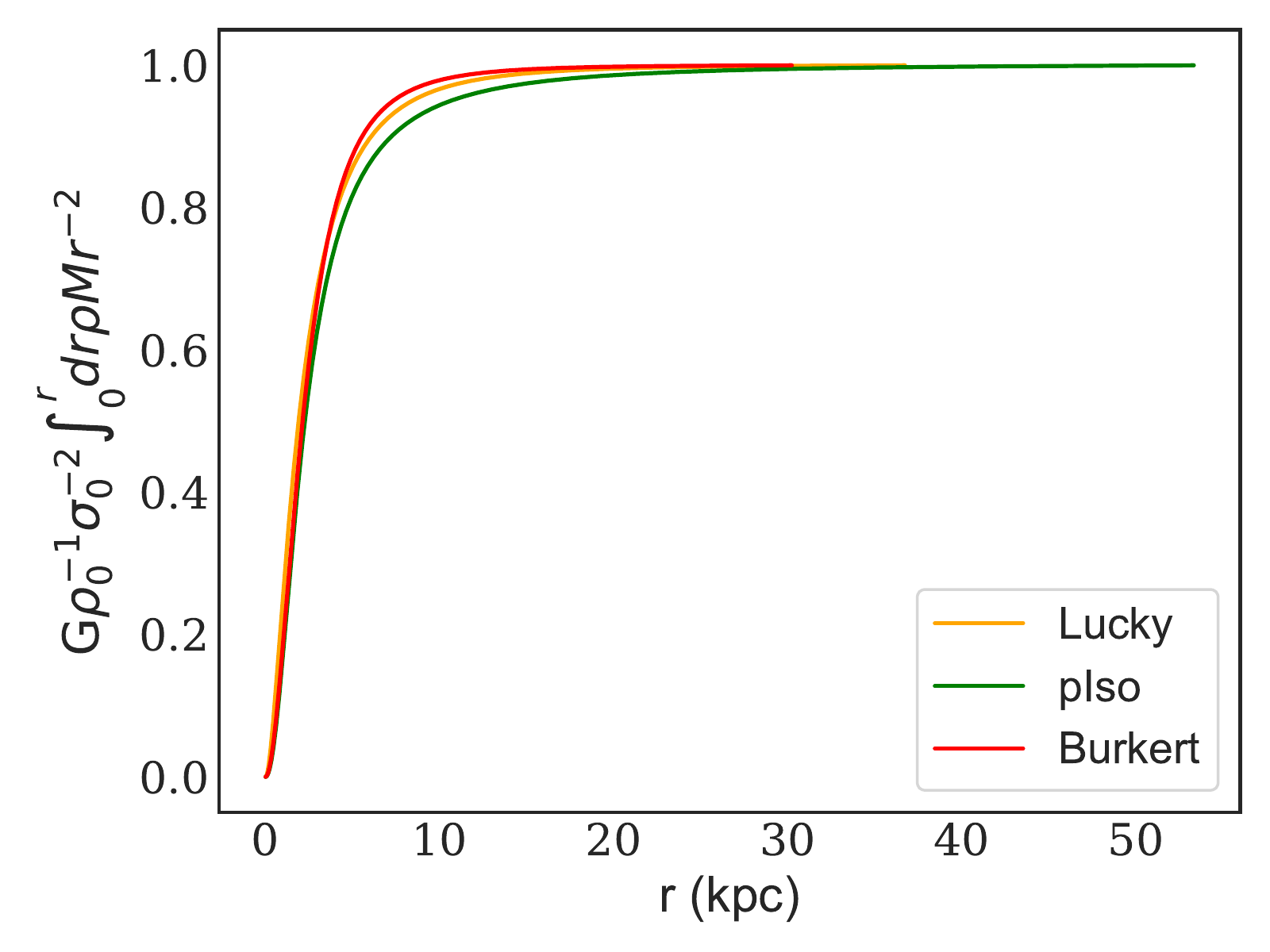}
\caption{The second term in the parentheses in equation~\eqref{Eq:y_r} is plotted versus the distance from the center. The figure shows that this term does not vary with distance in the outer region of halos. We report the same behavior in all of the galaxies although this plot refers particularly to DDO170. \lb{Fig:iGal_8_YDependenceOnR}}
\end{figure}
To investigate how $T_R$ varies with respect to our definition of the edge of the halo, we inspect the components of equation~\eqref{Eq:y_r}. 
As can be seen in figure~\ref{Fig:iGal_8_YDependenceOnR}, the second term in the parentheses reaches a plateau in the outer regions. Therefore, at large distances from the center, $y(r) \propto \rho(r)^{-1}$, and $T_{R_l} = 200 l^{-1}T_{R_{200}}$, where $R_l$ is a radius at which the mass density of the halo is $l$ times the critical density of the universe.  This dependency of temperature on density has been tested via a direct computation as well. 
Consequently, the mass of DM in terms of its temperature at the edge reads
\bqn
\lb{Eq:mOverT}
 m \simeq \frac{l}{200}\times \frac{10^6}{1.9} \times T(R_l) (\text{eV}\cdot \text{K}^{-1}).
\eqn

\begin{figure}
\centering
\includegraphics[width=\columnwidth]{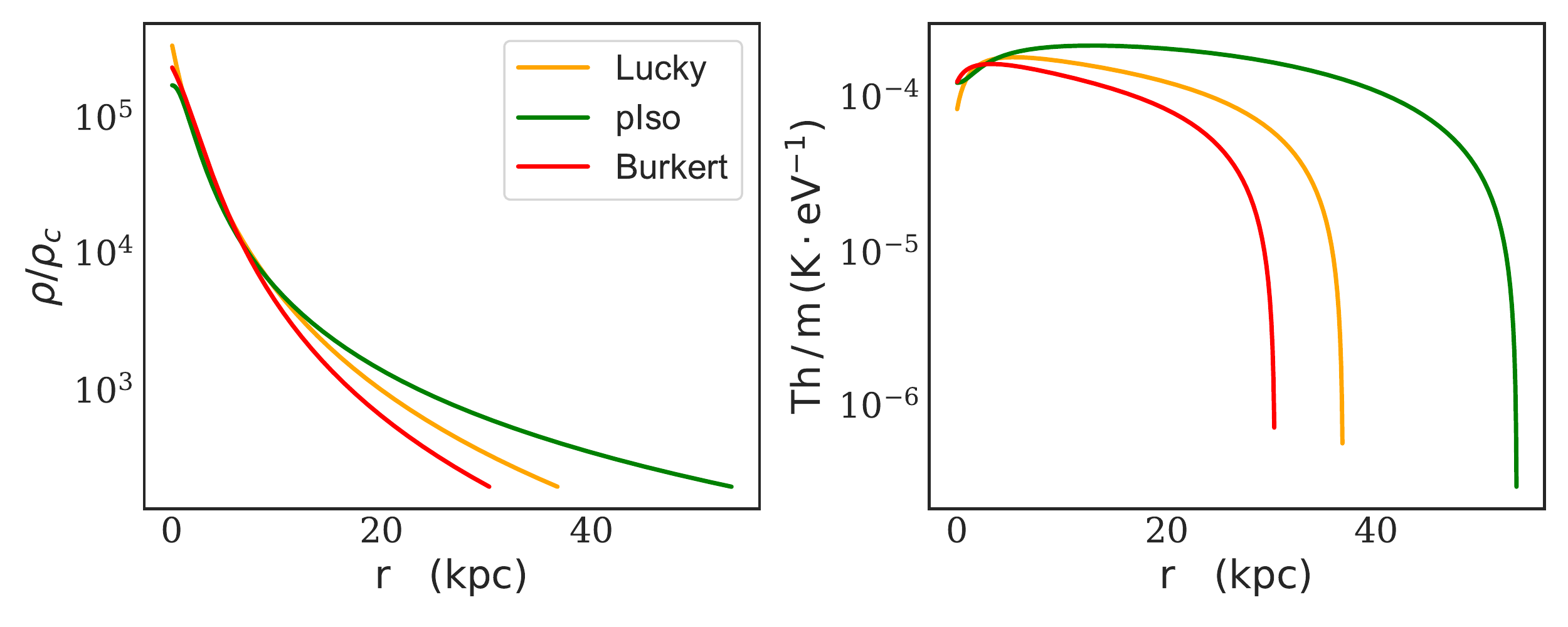}
\caption{The mass profile, and $\sigma^2(r)/k$ profile of the halo of DDO170 galaxy.  Assuming that at the edge $h_R\simeq 1$,  the edge temperature over the mass of DM is $\simeq 10^{-6}$(K/eV) consistent with figures~\ref{Fig:T_m_M_R200rhoc} and \ref{Fig:T_m_M_R200rhoc_LittleThings}. \lb{Fig:iGal_8_profiles}}
\end{figure}
Figure~\ref{Fig:iGal_8_profiles} shows the mass and temperature profiles of the halo of DDO170. It should be noted that the former is the observed mass model reported in \cite{2020ApJS..247...31L}. The latter is the combination of the observations and the theoretical framework presented in this paper.
As can be seen from the figure, the temperature increases with the distance from the center up to a point before it starts to decrease. 
This temperature profile can lead to an additional attractive force in the inner region of the halo. The origin of the extra attractive force can be understood by noting that the left hand side of equation~\eqref{Eq:Stability} is the force due to the pressure. Since both temperature and mass density are functions of $r$, the force of the pressure has two components. 
In figure~\ref{Fig:iGal_8_forces}, we show the two components of the force of pressure relative to gravity, where $\frac{dP_1}{dr}$ is due to  the temperature derivative. The figure illustrates that the extra attractive force can be orders of magnitude stronger than gravity at the center leading to higher compression of DM and lowering the so-called phase-space lower bounds on the mass of DM \cite{2020EPJC...80.1076B}.

\begin{figure*}
\centering
\includegraphics[width=\columnwidth]{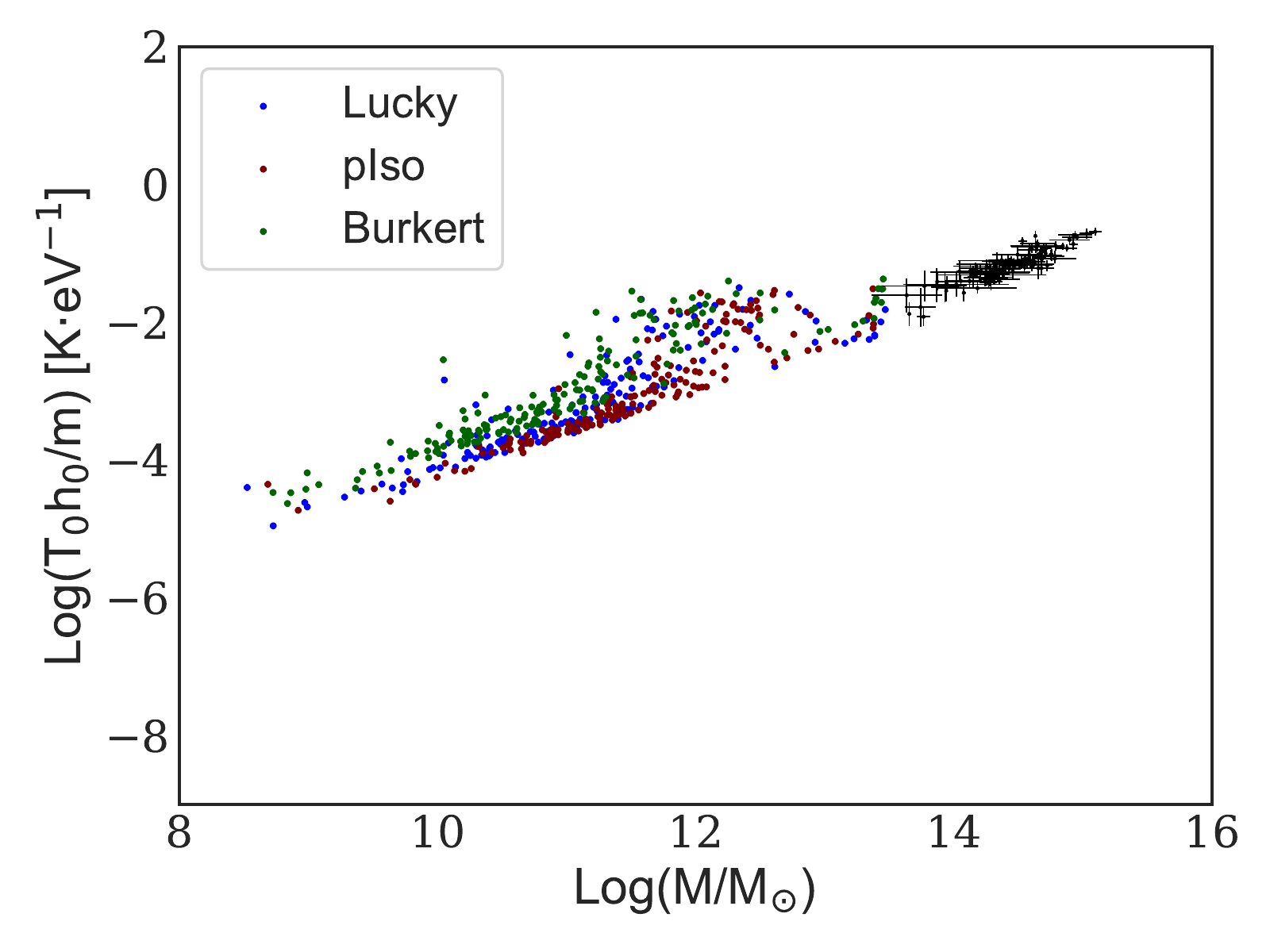}
\includegraphics[width=\columnwidth]{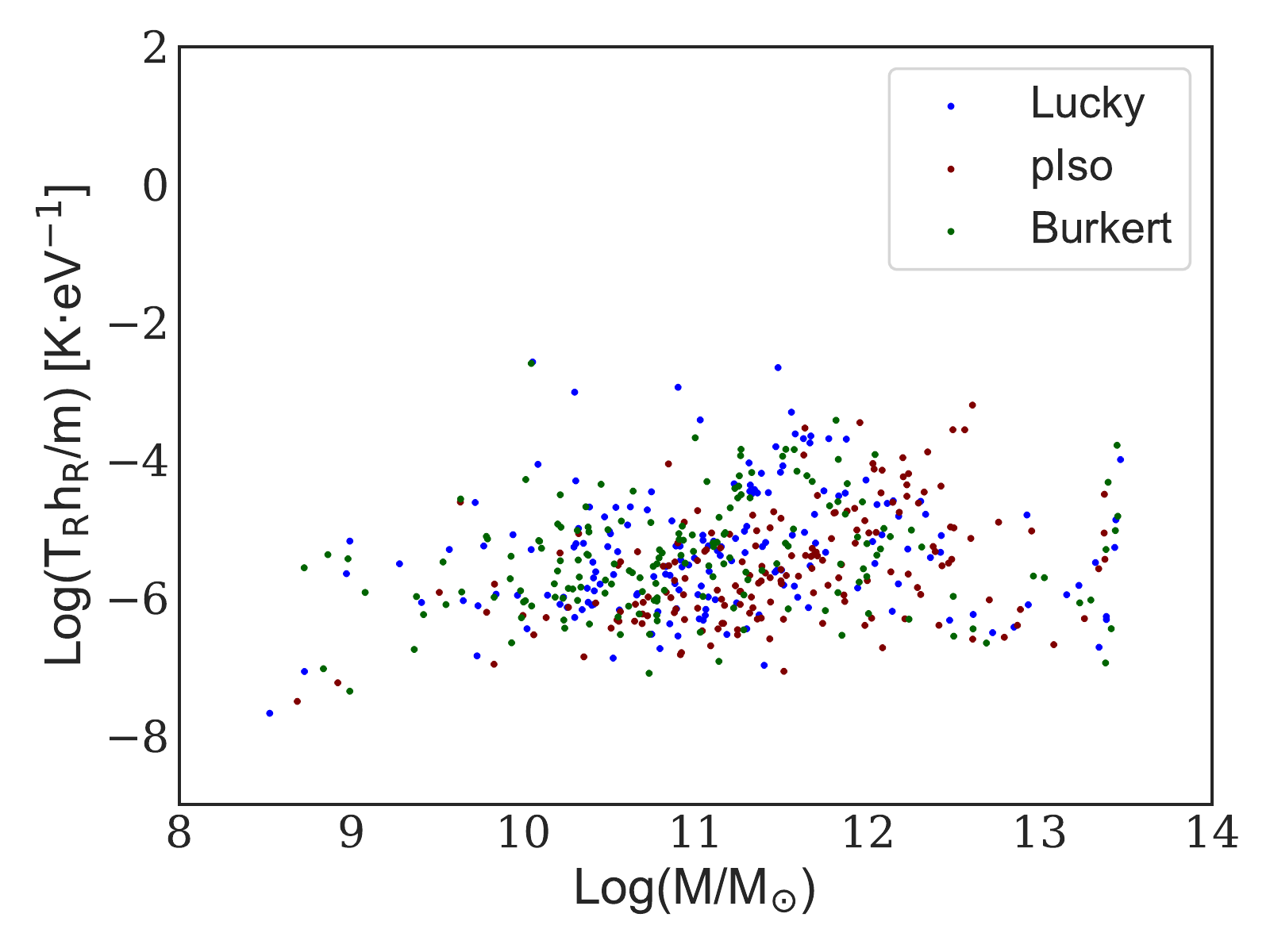}
\caption{The temperature to the mass of DM versus the total mass of the halos of the SPARC galaxies,  after taking the visible matter into consideration. The left panel refers to the DM temperature at the center of the halos and the right panel refers to the temperature of DM at the edge of the halos where the mass density of DM is only 200 time the critical mass density in the universe.  
The points with error bar in the left panel are from \cite{2016ApJ...819...63R} and show the observed ``temperature over mass'' of visible matter in Sunyaev-Zeldovich-selected clusters.  Temperature of DM at the center shows an overall increase due to the visible matter.  At the edge,  the temperature is more scattered than in the right panel of figure~\ref{Fig:T_m_M_R200rhoc} although still at 95\% confidence $T(R_{200})/m=\left(3.6 \pm 0.5\right)\times 10^{-6}\,(\text{K}\,\cdot\,\text{eV}^{-1})$. 
\lb{Fig:T_m_M_R200rhoc_withBaryons}}
\end{figure*}
\section{The effects of the visible matter}
\lb{Sec:VisibleMatter}
In this section,  we take the visible matter of the SPARC galaxies into consideration and show the changes in the estimations for their $\sigma_0^2$ and $m/T(R)$.  
The visible matter contribution to the rotation curves in the SPARC dataset is decomposed into a disk of stars,  a disk of gases,  and a spherical bulge,  and is presented in \cite{2016AJ....152..157L}.  We use the observations to fit the disks to exponential surface density profiles and the bulges to exponential volume density profiles.  
For the sake of simplicity,  it is possible to sphericize the disk contributions \cite{2020Univ....6..118S}.  For both of the disks, we assume that the gravitational force is equal to $-G M_{\text{disk}}(r)/r^2$ in all directions.  Given the small size of the disk in comparison to the size of the halo,  this should be a fairly good approximation.

In the presence of visible matter,  the gradient in the pressure of DM should confront the gravitational force due to the total enclosed masses.  Therefore, the mass profile in equation~\eqref{Eq:Stability} should be replaced by $M(r)\rightarrow M(r) + M^*(r)$,  where asterisk stands for visible matter and no asterisk refers to DM.  The dimensionless dispersion velocity squared and the pressure read
\bqn
\lb{Eq:y_P_withBaryons}
&&y^{_{\text{new}}}(r) =  y(r) -  \frac{G}{\rho(r)\sigma_0^2}\int_0^{r}\frac{\rho(r')M^*(r')}{r'^{2}}dr',\nb\\
&&~\nb\\
&&P^{{\text{new}}}(r) = \sigma_0^{_{\text{new}}2}  y^{_{\text{new}}}(r)  \rho(r).
\eqn

To find the modified dispersion velocity,  we need to account for the visible matter's contribution to the virial theorem through the external force that it applies to the halo.  We start from the basic virial theorem
\bqn
-2 \sum_i  \frac{3}{2}P^{{\text{new}}}_i = \sum_i \vec{F}_i \cdot \vec{r}_i,  
\eqn
where the subscript i runs over every mass interval $dm_i$ in the halo,  and $\vec{F}_i $ is the total force applied to $dm_i$.  After breaking the total force to the internal force from DM and external force from visible matter,  and in the continuum limit we have
\bqn
\lb{Eq:CentralDispersion_withVisible}
\sigma_0^{_{\text{new}}2}  =  \frac{-1}{2\pi \rho_0 R^3}\left(\frac{1}{2}W^{_{\text{new}}}+U^{_{\text{new}}}_2\right),
\eqn
where 
\bqn
&&W^{_{\text{new}}} =  W -4 \pi G \int_0^R dr\, r\, \rho(r) M^*(r)\nb\\
&&U^{_{\text{new}}}_2 = U_2 -
 6\pi G \int_0^R r^2 dr \int_0^{r}\frac{\rho(r')M^*(r')}{r'^{2}}dr'.
\eqn
Finally,  the mass to temperature of DM at the edge of the halo reads
\bqn
\lb{Eq:DMmass_withBaryons}
\frac{m}{T^{{\text{new}}}(R)} = \frac{k}{\sigma^{_{\text{new}}2}_0 \, y^{_{\text{new}}}(R)}.
\eqn

At this point,  we follow the same procedure as in section~\ref{sec:Sig2OfSPARC} and estimate $\sigma_0^{_{\text{new}}2}$ and $m/T^{{\text{new}}}(R)$ for every galaxy in the SPARC dataset.  Figure~\ref{Fig:T_m_M_R200rhoc_withBaryons}(left) shows that the central temperature of DM is increased due to the contribution from the visible matter.  More explicitly,  the ratio $\sigma_0^{_{\text{new}}2}/\sigma_0^2$ is presented in figure~\ref{Fig:Ratio_Sig20_Baryons},  which shows that the central dispersion velocity squared is increased by a factor in the range of 1.5 to 25.  On the other hand,  figure~\ref{Fig:T_m_M_R200rhoc_withBaryons}(right) indicates that the temperature of DM at the edge of halos does not change significantly.  
The temperature of DM at $R_{200}$ divided by its mass,  at 95\% confidence,  reads
\bqn
\lb{Eq:TR200_m_withBaryons}
\frac{T(R_{200})}{m}=\left(3.6 \pm 0.5\right)\times 10^{-6}\,(\text{K}\,\cdot\,\text{eV}^{-1}),
\eqn
which comparing to equation~\eqref{Eq:TR200_m_withBaryons} shows little change.

\section{Cosmological Scenarios}
\lb{sec:Cosmology}
In the preceding sections, we expressed the mass of DM in terms of its universal temperature in the outer part of halos. We discussed that the temperature should have been set in the early universe. Hence, we should be able to find the mass of DM in terms of the temperature of CMB.
In the following, we investigate the relation between the two within the context of a few cosmological schemes.

\subsection{Non-Relativistic Decoupling}
If DM is cold, its decoupling temperature and the universal temperature in the outer regions of halos are related by the scale-factor through $T_{R_{l}} \simeq T\f a\f^2$. 
On the other hand, the freeze-out temperature of DM in terms of the present temperature of CMB reads $T\f \simeq T_{_{\text{CMB}_0}} a\f^{-1}$. Therefore, using equation~\eqref{Eq:mOverT}, the mass of DM reads
\bqn
\lb{Eq:massInTf}
m  = \frac{l}{200} \times 10^6 \frac{T_{_{\text{CMB}_0}}}{T\f}\, (\text{eV}),
\eqn
where we have used $T_{_{\text{CMB}_0}} \simeq 1.9$~eV to account for the electron-positron annihilation in the early universe. 

Since DM is assumed cold, its mass has to be larger or equal to its freeze-out temperature in natural units. Therefore, equation~\eqref{Eq:massInTf} leads to the following lower bound on the mass of DM 
\bqn
\lb{Eq:NonRelLowerMass}
\sqrt{\frac{1.6\, l }{200}}  10\, \text{eV} < m.
\eqn

On the other hand, the decoupling of DM from CMB has to be before the photon last scattering at $\sim 0.3\,$eV. This inequality and equation~\eqref{Eq:massInTf} leave the following upper bound on the mass of DM
\bqn
m < \frac{l}{200}500\, \text{eV}.
\eqn

Since the mass density at the edge of halos cannot be smaller than the cosmological DM mass density, we expect that $0.3 < l < 200$.  Hence, the mass of CDM falls in the following range
\bqn
0.5\,\text{eV} < m < 500\,\text{eV}.
\eqn 

The largest value of the ratio of the Fermi-Dirac integrals in this case is equal to $h_R\simeq 1.7$,  see appendix~\ref{App:hR},  which is still reasonably close to 1.

Using equation~\ref{Eq:massInTf}, in this model, DM decouples from CMB when the temperature of the radiation is between 
$0.3\, \text{eV} < T\f < 10\,\text{eV}$. This scenario is not acceptable since otherwise, we should have seen the signature of DM in the particle physics experiments.  
Therefore,  CDM is at odds with the observations and is not a viable scheme.

\subsection{Relativistic Decoupling}
If DM decouples from the visible matter when it is still relativistic, the temperature at the edge of galaxies is approximately equal to the present temperature of CMB.  
Hence, using equation~\eqref{Eq:mOverT}, the mass of DM reads
\bqn
m \simeq \frac{l}{200}\times10^6 (\text{eV}). 
\eqn

Since we expect that $0.3 < l < 200$,  the equation above implies that the mass of DM is in the following range
\bqn
1.5\, \text{keV} < m < 1\, \text{MeV}.
\eqn
The estimated mass range is above the temperature of the matter-radiation equality $\sim 1\,$(eV), which subsequently means that DM is warm in this scenario.
We can also conclude that a hot dark matter model is not consistent with the analyzed observations of this paper.

Remarkably,  our estimation for the mass range of WDM is the same as the estimated mass range for sterile neutrinos as the most popular candidate for WDM \cite{2017JCAP...01..025A}.   
It is even more surprising that observation of rotation curves in galaxies is in favor of a DM model that corresponds with the only sector of the standard model of particle physics that needs revision at the moment.

\section{Conclusions}
\lb{sec:conclusion}
We have analyzed observations of around 200 late-type galaxies by assuming that (i) DM obeys either the Fermi-Dirac or the Maxwell-Boltzmann distribution, and (ii) the halos are in the virial state. 
Hence, using the stability of the halos, the temperature of DM has been expressed in terms of the mass density that is estimated from observations. 

The dispersion velocities of DM at the center of the halos have been estimated. We have shown that the dependence of the latter on the total mass of the corresponding halos is the same as in N-body simulations, suggesting that the halos are in the virial state as assumed. 

The temperature of DM at the center and the edge of the halos have been estimated. At the center, DM temperature increases with the total mass of the halo. However, at the edge, the temperature is independent of the halo properties and its ratio with respect to the mass of DM is universally equal to $\left(1.9\pm 0.3\right)\times 10^{-6}$~(eV$\,\cdot\,$K$^{-1}$) at 95\% confidence.  
This result indicates that DM at the edge of the halos is an unperturbed thermal relic whose temperature is equal to its decoupling temperature in the early universe, times a power of the scale factor due to the expansion of the universe. 
It should be noted that the universality of the temperature of DM in unperturbed regions of the universe is the natural consequence in any thermal model of DM.  However, we do not impose such universality by assumption. Instead, we derive it from observations.

We have studied the implications of the observed universal temperature within the context of two cosmological models. It has been shown that hot DM, and CDM are not consistent with the observations, while WDM scenario with a mass of 
$1.5\,$(keV)$\,<m<\,1\,$(MeV) is a viable scenario.

\appendix
\section{Estimation of $h_R$}
\lb{App:hR}
In equation~\eqref{Eq:DMmass}, we assumed that the ratio of Fermi-Dirac integrals is equal to one at the edge of the halos $h_R \simeq 1$, meaning that DM obeys the Maxwell-Boltzmann distribution at the edge of halos despite its fermionic nature. 
However, the assumption is redundant and could have been avoided. In this section, we explicitly calculate $h_R$ for the scenarios reported in section~\ref{sec:Cosmology}. 

First, we define $X \equiv m^{\frac{5}{2}} T^{\frac{3}{2}}$. From statistical mechanics, it can be expressed in terms of the mass density, and the fugacity of fermionic DM 
\bqn
\lb{Eq:XR2}
X_R = \frac{\rho_R}{2^{\frac{5}{2}}\pi^{\frac{3}{2}} f_{\frac{3}{2}}(z_R)},
\eqn
where the subscript $R$ is because we are interested in the quantities at the edge of the halos. 

In the relativistic freeze-out scenario and the upper-mass in the non-relativistic decoupling scenario of section~\ref{sec:Cosmology}, the temperature $T_R$ is directly known in terms of the temperature of CMB. Hence, we rewrite equation~\eqref{Eq:mOverT} into
\bqn
X_R = \left( \frac{l}{200*1.9*1.6}\right)^{\frac{5}{2}} 10^{25}\, T_R^4\, h_R^{\frac{5}{2}}.
\eqn

To find the solution to the fugacity $z_R$, we plot the two $X_R$ above for $z_R=0, \cdots, \infty$ and observe where they cross.  In both of the mentioned scenarios $z_R \ll 1$, which subsequently means $h_R \simeq 1$. 

In the lower-mass in the non-relativistic decoupling scenario of section~\ref{sec:Cosmology}, the temperature $T_R$ is given in terms of the CMB temperature and the mass of DM. Hence, we can rewrite equation~\eqref{Eq:mOverT} into
\bqn
X_R = \left( \frac{l}{200*1.9*1.6}\right)^{\frac{1}{2}} 10^{5}\, T_{\text{CMB}_0}^4\, h_R^{\frac{1}{2}}.
\eqn

Again, we plot this $X_R$ and the one in equation~\eqref{Eq:XR2} for $z_R=0, \cdots, \infty$. The crossing point of the two curves would be the solution to $z_R$.  
The largest value of the fugacity in this case is  equal to $z_R\simeq 20$,  which shows deviation from classical distribution. However,  the ratio of the Fermi-Dirac integrals for the latter is only $h_R \simeq 1.7$ which is still close to what we assumed in equation~\eqref{Eq:DMmass}.

\section{Additional figures}
\begin{figure}[b]
\centering
\includegraphics[width=\columnwidth]{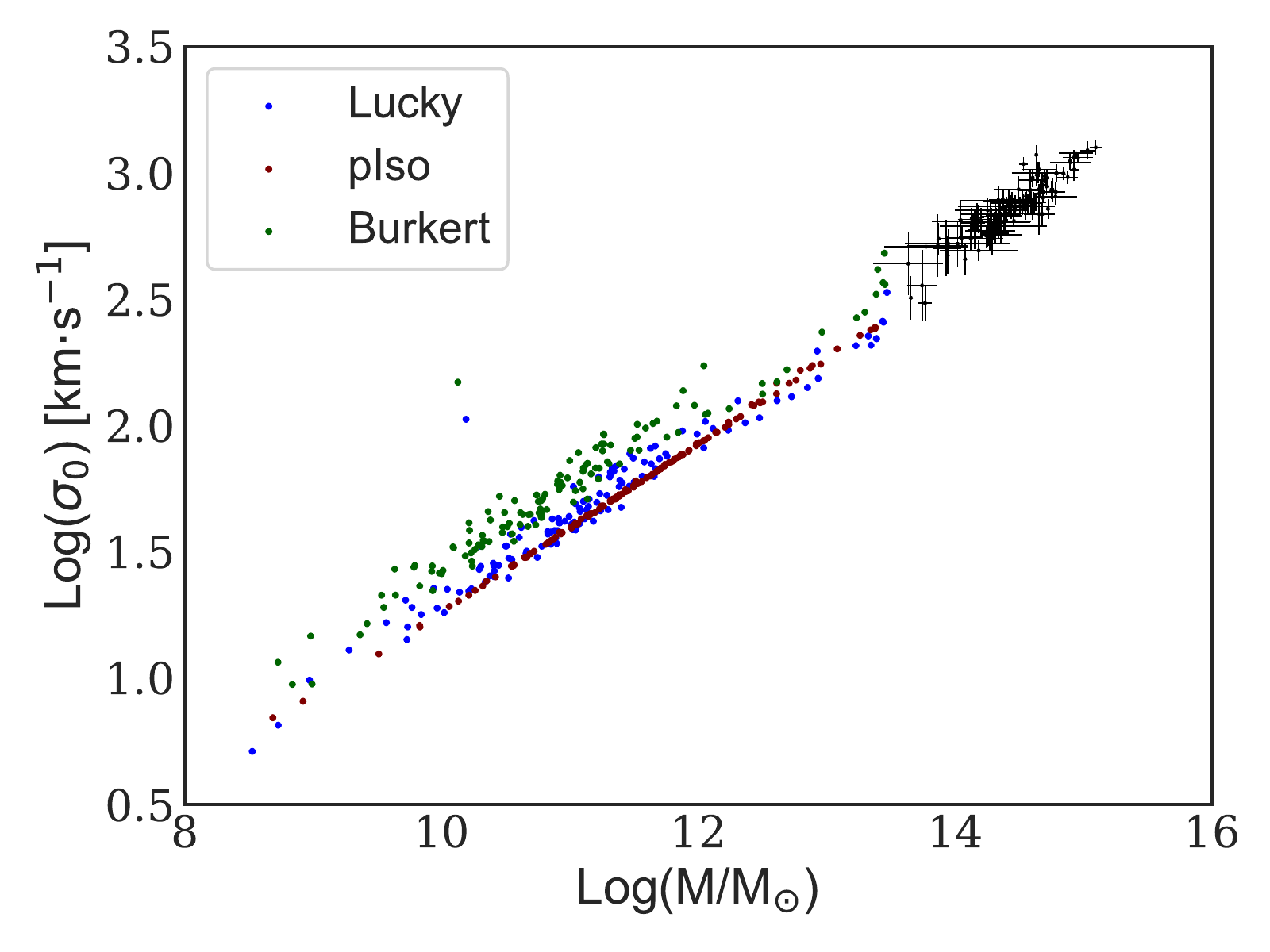}
\caption{The central dispersion velocity of DM as a function of the total mass of the halo for the SPARC galaxies.  It should be noted that this figure is nearly insensitive to where the edge of the halo is defined. The points with error bar are from \cite{2016ApJ...819...63R} and show the observed dispersion velocity of visible matter in Sunyaev-Zeldovich-selected clusters. \lb{Fig:Sig0_M_R200rhoc}}
\end{figure}

\begin{figure*}[b]
\centering
\includegraphics[width=\textwidth]{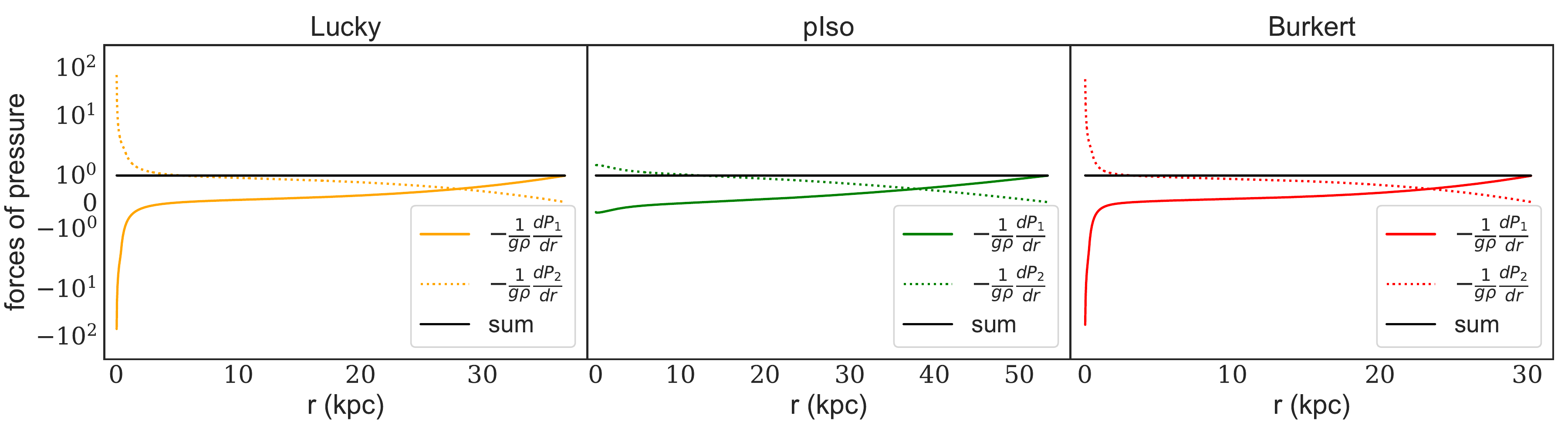}
\caption{Each of the two components of the force of the pressure is divided by the gravitational acceleration. The component containing the temperature derivative is indicated by $\frac{dP_1}{dr}$. As can be seen, the forces of the pressure are significantly larger than the gravitational force at close to the center of the halo. Since the sum of the two ratios is equal to one at all distances, the gravitational force is strong enough to balance the repulsive force and maintain the stability.  \lb{Fig:iGal_8_forces}}
\end{figure*}

\begin{figure}[b]
\centering
\includegraphics[width=\columnwidth]{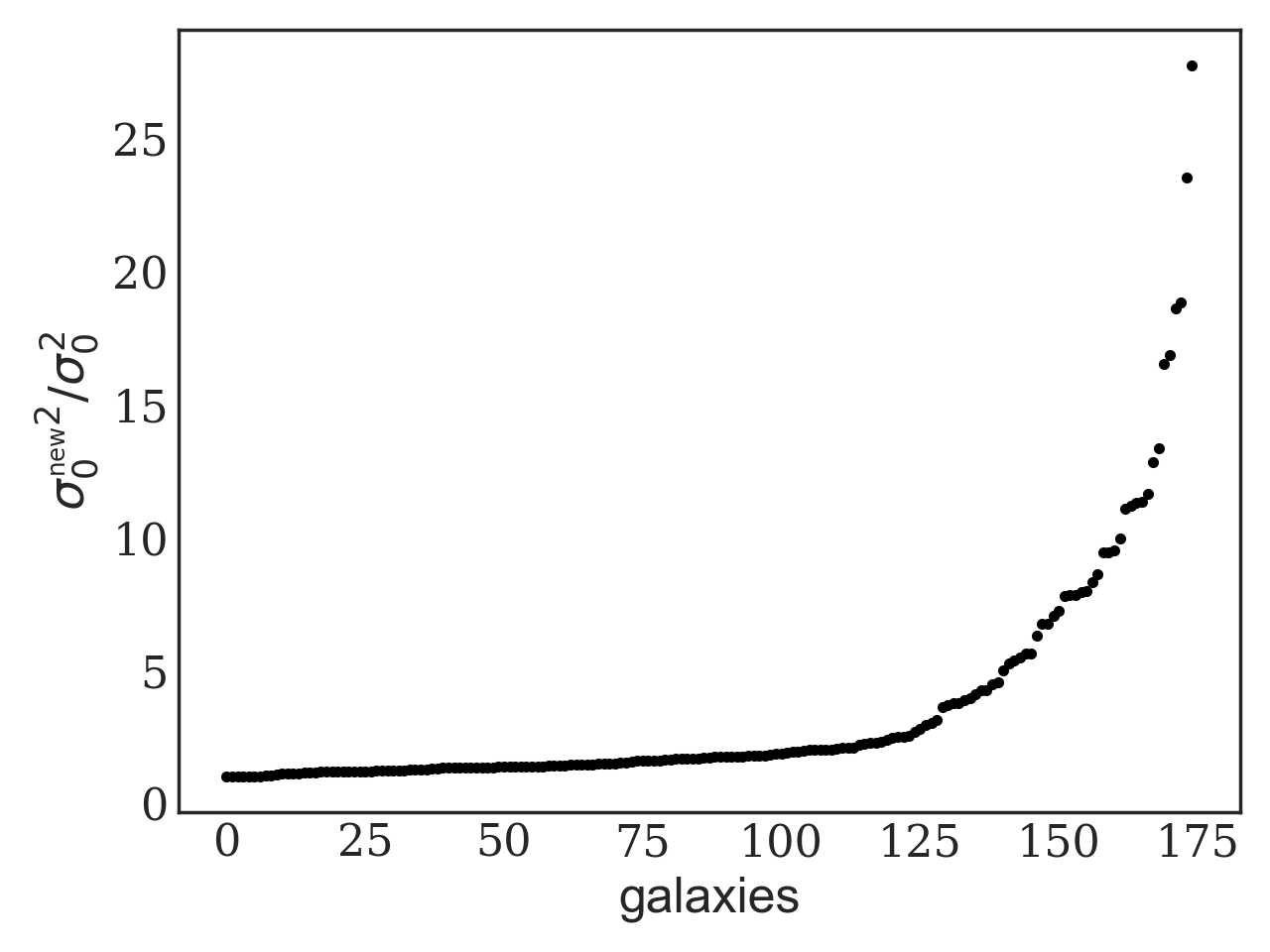}
\caption{
Ratio of the dispersion velocities at the center $\sigma_0^{_{\text{new}}2}/\sigma_0^2$ for 175 galaxies in the SPARC database.  Visible matter increase the central dispersion velocity in all galaxies by a factor ranging from 1.5 to 25.  The galaxies are sorted in the x-axis based on their corresponding ratios in the y-axis.  \lb{Fig:Ratio_Sig20_Baryons}
}
\end{figure}


\bibliography{Refs}

\end{document}